# Atomically resolved phase coexistence in VO$_2$ thin films

Masoud Ahmadi[1]\*, Atul Atul[1]\*, Sytze de Graaf[1], Ewout van der Veer[1], Ansgar Meise[2], Amir Hossein Tavabi[2], Marc Heggen[2], Rafal E. Dunin-Borkowski[2], Majid Ahmadi[1], Bart J. Kooi[1]\*

[1]*Zernike Institute for Advanced Materials, University of Groningen, Nijenborgh 4, 9747, AG, Groningen, the Netherlands*

[2]*Ernst Ruska-Centre for Microscopy and Spectroscopy with Electrons (ER-C), Forschungszentrum Jülich, 52425 Jülich, Germany*

\*Corresponding authors: masoudahmadi777@gmail.com, atul061094@gmail.com and b.j.kooi@rug.nl

**Abstract**

Concurrent structural and electronic transformations in VO$_2$ thin films are of twofold importance: enabling fine-tuning of the emergent electrical properties in functional devices, yet creating an intricate interfacial domain structure of transitional phases. Despite the importance of understanding the structure of VO$_2$ thin films, a detailed real space atomic structure analysis in which also the oxygen atomic columns are resolved is lacking. Moreover, intermediate atomic structures have remained elusive due to the lack of robust atomically resolved quantitative analysis. Here, we directly resolve both V and O atomic columns and discover the presence of the strain-stabilized intermediate monoclinic (M2) phase nanolayers (less than 2 nm thick) in epitaxially grown VO$_2$ films on a TiO$_2$ (001) substrate, where the dominant part of VO$_2$ undergoes a transition from the tetragonal (rutile) phase to the monoclinic M1 phase. We unfold the crucial role of imaging the spatial configurations of the oxygen anions (in addition to V cations) utilizing atomic-resolution electron microscopy. Our approach sets a direct pathway to unravel the structural transitions in wide range of correlated oxides, offering substantial implications for e.g. optoelectronics and ferroelectrics.

The versatility and exceptional properties of $VO_2$ thin films have enabled the emergence of tunable functional properties for modern switching devices, neuromorphic sensors and optoelectronics [1–4]. This is predominantly due to the reversible metal to insulator transition (MIT) in this correlated oxide, which occurs slightly above room temperature (~ 68 °C) from metallic rutile structure (at higher temperature) to monoclinic insulator phases (at lower temperature) [5,6]. Nevertheless, the exact mechanism of the MIT with concomitant structural/electrical transformations in this scheme has remained debatable so far. From the structural point of view, an ambiguity has arisen, not only because of the complex interfacial domains, but also from the lack of atomically resolved quantitative information by means of direct characterization techniques.

The MIT scheme of $VO_2$ in thin films, nanowires and bulk formats has been a subject of a long-lasting study with different aspects of attention [7–25]. One of the less-understood features in these films are the structural alterations at the atomic scale that occur during the transition. These films undergo a martensitic-like first order transition from high temperature tetragonal metallic phase i.e. rutile R phase with space group: *$P4_2/mnm$*, to a low temperature insulating monoclinic phase i.e. M1 with space group: *$P2_1C$* [26]. In addition to these typical phases, some metastable and transitional phases including M2, T and xM3 have been also reported [27–31]. Mediated by strain, a complex phase coexistence in these films has been identified [32]. Among the phases present, the monoclinic M2 (space group: *C2/m*) has received particular attention as it provides extensive opportunities (superior to M1) for electronic properties modulation, novel memory applications and thermal actuators [33,34]. Nonetheless, the spatial foundation and atomic structure characteristics of the M2 phase and its relation to the MIT require further understanding.

So far, the transitional domains during the MIT in $VO_2$ have remained unresolved at atomic scale due to their complexity. Intermediate phases (specially the M2 polymorph) have been either predicted theoretically using e.g. density functional theory (DFT) calculations [35] or have been spotted by Raman spectroscopy [36], X-ray diffraction [32] and conductive atomic force microscopy [37]. However, these methods offer characteristic information on a more global scale and lack the local atomic-scale detection of these phases with exact spatial coordination (distribution), particularly in thin films. Naturally, atomic scale characterization would provide invaluable information on the MIT, as it initiates at this scale, and can enable tunable tailored-made functional materials and switching devices. Imaging the oxygen atoms (V-O bonds) in these films with aberration-corrected scanning transmission electron microscopy (STEM) can reveal atomic-scale electrical or structural changes enabled by oxygen atoms configuration or displacement. In this context, the approach we present here is believed to open an avenue for quantifying the local atomic structures in a wide range of correlated oxides.

The basis of our work is the successful growth of high-quality $VO_2$ thin films on $TiO_2$ (001) substrates using pulsed laser deposition (Fig. S1), schematically shown in Fig. 1A. Only for this substrate and specific surface orientation the $VO_2$ atomic structure can be unambiguously resolved in projected images, for which we also provide more background information in another paper [38]. We start here by schematically revealing the different structures in the specifically chosen crystallographic projections that play a key role in our approach to properly resolve all the atomic structure details in our $VO_2$ films. Overview scanning electron microscopy image and atomic force microscopy map of the $VO_2$ domains are shown in Fig. S2. As Fig. 1 depicts, the $VO_2$ grown at a temperature of 400 °C undergoes a martensitic-like transformation (at about 68 °C) from the rutile R phase (panel B) to the intermediate phase M2 (panels C and D) reaching the M1 phase (panels E and F) at room temperature. The $VO_2$ film is epitaxially grown on the (001) surface of $TiO_2$. Then, when the $VO_2$ is imaged in cross-sections with the interface edge-on, the M2 and M1 phases will be observed along two orthogonal crystallographic projection directions. M2 (A) type refers to the projection along $TiO_2$ $[100]_R$ and the M2 (B) along $TiO_2$ $[010]_R$ as depicted in Fig. 1C and Fig. 1D, respectively. M1 (B) and M1 (A) denote the projections along $[010]_R$ and $[\bar{1}00]_R$, respectively. In Fig. 1G and Fig. 1H, we introduce our criteria to distinguish the different phases. In Fig. 1G, the archetypal $VO_6$ octahedron for both $TiO_2$ ($VO_2$) tetragonal R phase and monoclinic M phases is given. Note that, apart from a small lattice parameter change, the $VO_6$ octahedron is identical for the R phases of $TiO_2$ and $VO_2$ (where the latter is only stable above about 68 °C). The vanadium atom is positioned in the center of the octahedron surrounded by six oxygen atoms labeled from O1 to O6, where each O-atom possesses a specific distance from the central V (i.e. V-O bond distance) and a certain angle in different configurations during the MIT. Table S1 provides the calculated theoretical values of bond distances and angles for the R and M structures. We also introduce the concept of ellipticity of the atoms in Fig. 1G, indicating the aspect ratio of the projected atoms in a column in an imaging zone axis. The ellipticity shows not only a size, but also a directionality, which is a useful criterion to be verified experimentally. Fig. 1H then shows a table where the different criteria, including atomic ellipticity, symmetry of V-O bonds with respect to central V atom and dimerization of the (edge-on observed) V-V planes, are applied to differentiate the structures in their different projection directions. Indeed, this table demonstrates that it is possible to distinguish the R, M2 and M1 phases in their different projections unambiguously.

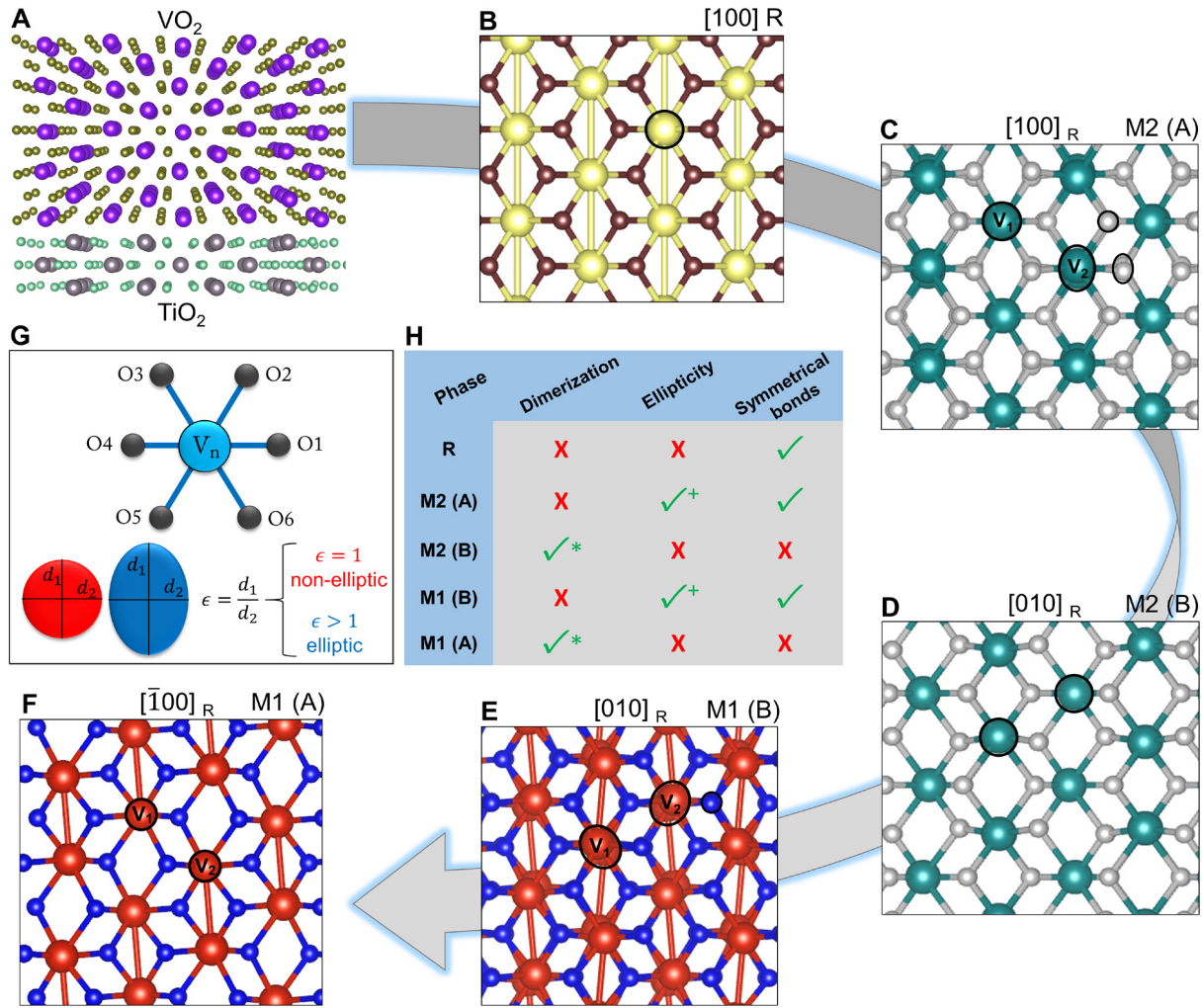

**Fig. 1. The crystallographic representations of observed structures in the MIT scheme and the criteria to distinguish these structures.**

The crystallographic representations of (A) the epitaxially grown $VO_2$ on a $TiO_2$ (001) substrate (B) $TiO_2$ rutile (R) phase in $[100]_R$ projection, (C) intermediate insulating monoclinic M2 (A) phase in $[100]_R$ projection, (D) intermediate insulating monoclinic M2 (B) phase in $[010]_R$ projection, (E) insulating monoclinic M1 (B) phase in $[010]_R$ crystal projection, (F) insulating monoclinic M1 (A) phase in $[\bar{1}00]_R$ projection, (G) the archetypal $VO_6$ octahedron indicating the labeled atoms together with the ellipticity concept, (H) incorporated criteria to differentiate the present phases. + Note that M2 (A) and M1 (B) both generate clear ellipticity in their projections, yet the ellipticity of V-atoms in M2 (A) only occurs every second atomic columns (i.e. in V2), whereas the ellipticity of M1 (B) occurs for all the V-atomic columns. * Notice that the dimerization of (edge-on observed) V-V planes in M1 (A) is more pronounced than that of M2 (B).

We continue by showing the high-resolution STEM results of the studied $VO_2$ thin film. Fig. 2 depicts experimental STEM images of a cross-section of the film, the corresponding simulated STEM images and post-processing quantification of atomic structures in both experimental and simulated images. A detailed description of the focused ion beam lamella preparation and STEM imaging are provided in the supplementary information. In order to quantify the V-O bond coordinations and lengths in a statistical manner, we have developed a MATLAB code specifically for this purpose (see supplementary information).

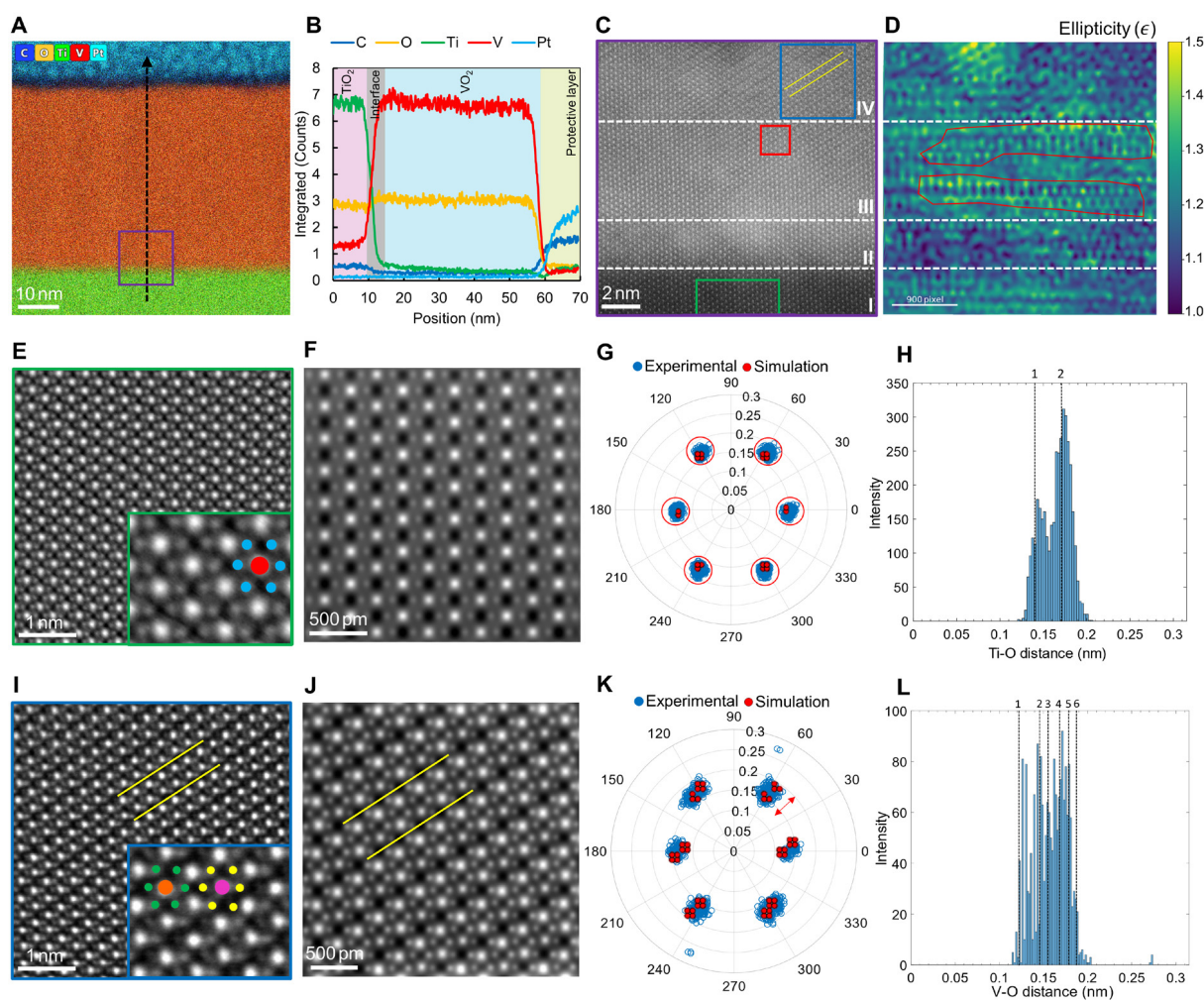

**Fig. 2. Quantitatively resolving the oxygen atoms in the rutile R and the dimerized M1 phase.**

(A) EDS map of a cross section of the film (in FIB-cut TEM lamella), (B) EDS elemental profiles along the indicated arrow in panel (A), (C) high-angle annular dark field (HAADF) image of the various domains of $VO_2$ film on $TiO_2$, (D) atomic ellipticity analysis on the entire region, M2 phase domains are designated by red contours, (E) integrated differential phase contrast (iDPC) image of R phase, (F) simulated iDPC micrograph of R phase, (G) quantified spatial configurations of the oxygen atoms for R showing symmetrical bonds (the units for angles and distances are ° and nm), (H) distribution of the Ti-O bond distances compared to theoretical values shown as 1 and 2 dashed lines, (I) iDPC image of the dimerized M1 (A) phase, (J) simulated iDPC image of the dimerized M1, (K) quantified spatial configurations of the oxygen atoms for M1 (A) revealing non-symmetrical bonds, (L) distribution of the V-O bond distances compared to theoretical values indicated as 1 to 6 dashed lines (12 measured bonds are grouped into 6 pairs as described in the supplementary information).

As can be observed in the EDX results in Figs. 2A and 2B, the representative selected area in the cross section consists of the $VO_2$ film on top of the $TiO_2$ substrate also including the interface (with thin intermixed region). However, the experimental atomic resolution HAADF-STEM image in Fig. 2C provides more details on the film. Four different zones are resolved, namely $TiO_2$ substrate (Zone I), intermixing region (Zone II), $VO_2$ region containing transitional domains M2 (Zone III) and the $VO_2$ region fully composed of M1 (Zone IV). The corresponding ellipticity map of the selected area is depicted in Fig. 2D, where particularly in Zone III distinct periodic patterns can be observed.

The iDPC-STEM image of the $TiO_2$ (001) substrate viewed along $[100]_R$ in Fig. 2E shows perfectly symmetrical Ti-O bonds, where oxygen atoms are distinctly resolved in accordance with Fig. 1B. The simulated iDPC-STEM image of the same zone axis in Fig. 2F confirms the revealed structure of the R phase. The quantification of the Ti-O bond distances and the spatial configurations of the oxygen atoms with the corresponding angles for the R phase are provided in Fig. 2G. The simulation results show that the oxygen atoms are accurately imaged in the zone axis with a precise spatial configuration in accordance with the R-phase model. We statistically confirm and measure the coordination of the O-atoms, which corresponds well with the theoretical calculations given in Table S1. To further elucidate the structure, the theoretical values of Ti-O bonds distances are compared (via dashed lines) to the measured experimental ones in Fig. 2H. Indeed, the two short and four long Ti-O bond distances are distinctively resolved resulting in two different peaks with the correct distances and intensity ratio of one-to-two in the histogram.

The experimental and simulated iDPC-STEM images of the $VO_2$ M1 (A) phase (from Zone IV) are shown in Fig. 1I and Fig. 1J, respectively. Interestingly, these images clearly reveal for the first time the V-V dimerized atoms (the dimerization direction is defined by the tilted yellow lines) together with the oxygen atoms. The resolved structure is in accord with the M1 (A) crystal configuration (Fig. 1F) that exhibits two types of vanadium atomic columns: V1 and V2. Furthermore, the visualized quantification of oxygen atom positions using the iDPC images in Fig. 1K provides insightful observations. First, one can notice that, in contrast to the R-phase, the V-O bonds are not symmetrical anymore in this M1 phase. Second, the O-atoms possess a certain directionality relative to the dimerization direction. The simulated results manifest that there exist twelve calculated V-O bond types in the M1 (B) phase (see also Table S1). The plot in Fig. 2L elucidates the values of the V-O bonds in comparison with theoretical values. Note that the twelve theoretical values of V-O bonds from Table S1 are grouped in six pairs. As it can be conceived from Fig. 2L, in contrast to the distinct two peaks observed for the R-phase, here obviously a larger variety of V-O bond peaks occurs. Therefore, the atomic-structure electron microscopy analysis clearly distinguishes the R phase and M1 (A) in accordance with the criteria that were set earlier in Fig. 1H, i.e. a phase which does not exhibit dimerization and ellipticity but has symmetrical bonding is necessarily the R phase, whereas the $VO_2$ phase which does not possess ellipticity and symmetrical bonds, but exhibits dimerization of V-V atoms is indeed the M1 (A) polymorph.

Next, we proceed with the elaboration on the M1-M2 phase coexistence in the epitaxially grown $VO_2$ thin films. As it was shown in Fig. 2D, an inhomogeneous atomic ellipticity throughout the cross-section of the film is found. Surprisingly, right in Zone III, one can notice in certain regions a distinct pattern in the ellipticity map as indicated by the red contours in Fig. 2D. Although the imaging is performed completely in the zone axis for the $TiO_2$ substrate below zone III and the $VO_2$ M1 phase above zone III, atomic columns in a large fraction of zone III show substantial ellipticity in an alternating pattern parallel to the edge-on interface. Accordingly, the iDPC image is obtained from the selected area designated by the red square in Fig. 2C (from the same region where the distinct ellipticity pattern is observed) and presented in Fig. 3A. Thorough post-processing image analysis is applied to this image to unravel its reference structure. In Fig. 3A, one can observe that in (~2 nm) thin layers, some atomic columns are clearly elliptic, while the other alternating ones in a direction parallel to the interface are more localized and round. We now refer to specific structures already shown in Fig. 1, where M2 (A) polymorph exhibits exactly this ellipticity pattern for only every second V atomic column parallel to the interface in $[100]_R$ projection. Accordingly, the simulated iDPC image associated with M2 (A) phase matches well the experimental image in panel B of Fig. 3. The quantifications of the structures from both experiment and simulation (Fig. 3C) converge well and indicate the symmetrical bond distribution. In addition, the two distinct peaks appearing in the histogram (Fig. 3D) further confirm the symmetrical bonds in accordance with the theoretical values of the projected M2 (A) structure. As Fig. 3E and Fig. 3F specify, the blue atomic column (i.e. V1) uniformly shows a half-peak intensity value of 107 pm in out-of-plane direction across the atomic column diameter, whereas the red atomic column (V2) exhibits a value of 146 pm in this direction. This evidently implies that clear ellipticity occurs for the V2 atomic columns, whereas it is absent for the V1 atomic columns. It is interesting to note that, as it was found in Fig. 1C, the lateral oxygen atoms in $[100]_R$ projection of M2 (i.e. M2 (A)) also display a specific pattern of directional ellipticity. This subtle point is confirmed by experimental observation given in Fig. 3F, where for the O1 atomic column the intensity length (78 pm) is considerably larger in out-of-plane direction than that of the O2 (60 pm). These quantifications with their specific directionalities strongly confirm the coexistence of the intermediate M2 phase in a few nanometer thick layers of the $VO_2$ film. In other words, the transitional domain (i.e. Zone III) right above the intermixing region is dominated by the intermediate M2 (A) phase (see also Fig. S6). In particular, a polymorph that exhibits ellipticity every second V atomic column (and also ellipticity in specific O atomic columns), and possesses symmetrical bonds (two V-O distance peaks) with no V-V dimerization is indeed the M2 monoclinic phase discovered here. Note that we have not identified the other expected orientation of M2 (i.e. M2 (B)) by the STEM imaging, yet, for completeness we have included this orientation in our framework (Fig. 1) to clarify that also this projection direction must be present.

Figs. 3G-L illustrate the images associated with a region of the $VO_2$ film where both V-atomic columns show tilted ellipticity (see Fig. 3K) and oxygen atoms are completely circular (see Fig. 3L). The agreement between simulated and experimental quantification results demonstrates the presence of relatively symmetrical V-O bonds. These observations fit well with another orientation of M1, namely M1 (B) in $[010]_R$ projection. In contrast to previous phases, the strong V-atom ellipticity in M1 (B) results in a wide spread of the collected data points. In other words, the larger the atomic ellipticity, the less precisely defined the atomic center of mass becomes during image analysis, ultimately leading to more scattered O-atom positions. Nevertheless, as Fig. 3J demonstrates, in spite of this scattering, the two distinct peaks confirm the symmetrical bonds of the M1 (B) phase orientation.

In addition to the local atomic-scale investigations offered above, XRD reciprocal space mapping hints at the formation of the intermediate phases (in particular M2) within the $VO_2$ thin films on TiO$_2$ (001) substrate in a more global picture: see Fig. S7. The intermixing region (Zone II) in Fig. 1C and the long tails revealed in the reciprocal space map (RSM) in Fig. S7A, suggest the formation of some other intermediate phase(s) (e.g. T polymorph) as well [32]. However, this topic should be scrutinized in a separate focused study. The double stretched tails observed in the RSM of VO$_2$ [002] reflection (seen Fig. S7A) demonstrate a pronounced strain gradient in the film as was also suggested by Rodriguez et al. [32]. The strain state and its relationship with the transitory domain of the film are scrutinized next in a novel approach.

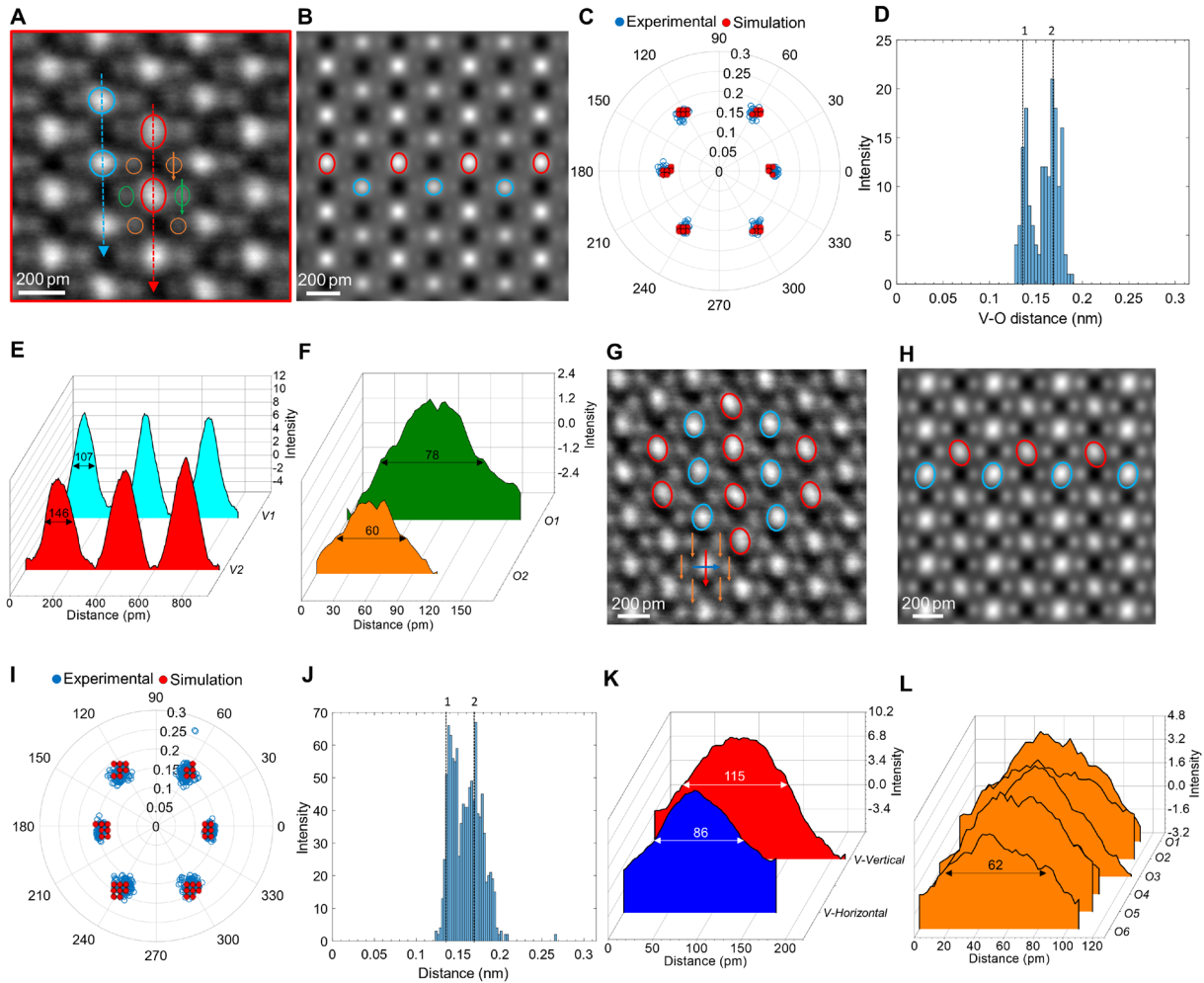

**Fig. 3. Real-space imaging of transitional phase.**

(A) Experimental iDPC image of the detected intermediate M2 (A) phase, (B) simulated iDPC micrograph of the M2 (A), (C) quantified spatial configurations of the oxygen atoms for M2 (A) (the units for angles and distances are ° and nm), (D) distribution of the V-O bond distances compared to theoretical values shown as 1 and 2 dashed lines, (E) intensity profiles along the V-atoms indicated in panel (A) demonstrating the alternating ellipticity in the V-columns, (F) intensity profiles along the O-atoms indicated in panel (A) showing ellipticity in O1, (G) iDPC image revealing M1 (B) phase, (H) simulated iDPC image of M1 (B) phase, (I) quantified spatial configurations of oxygen atoms for M1 (B), (J) distribution of the V-O bond distances compared to theoretical values shown as 1 and 2 dashed lines, (K) intensity profiles on V-atom indicated on panel (G) showing atomic ellipticity, (L) intensity profiles along O-atoms designated in panel (G).

Up to this point, the coexistence of the various $VO_2$ polymorphs, in particular the M2 phase has been resolved directly by atomic-structure electron microscopy observations. Nevertheless, it is required to get some insights into the potential physical origins of phase coexistence during MIT, specifically the stabilization of the intermediate M2 phase. It has been suggested in literature that the intermediate phases can be stabilized by strain, doping and electronic excitation [39–41]. Nevertheless, due to the lack of atomically resolved information, it has been a persistent challenge to correlate strain *locally* to the domains of the structural polymorphs. We aim to address this by mapping the strain in cross sections along the growth direction (c-axis, out-of-plane) and the in-plane direction of the film. To this purpose, atomic scale strain measurements using *real space* STEM images are accomplished. Fig. 4A and Fig. 4B display the strain distributions normal and parallel to the (edge-on) interface. These results cover Zones I, II, III right below the fully dimerized M1 phase region depicted in Fig. 2C. The corresponding iDPC image for the strain analysis is provided in Fig. S8. As it can be perceived from Fig. 4A-B, the absolute strain values in the film are compressive and tensile in the growth direction and in-plane direction, respectively. In particular in the growth direction (y-axis), one can notice a strong strain gradient, compressive in nature. Indeed, the Zone I ($TiO_2$ (001) substrate) as the reference exhibits almost zero strain value. As the strain profile in Fig. 4 denotes, by moving from the substrate to the film in the out-of-plane direction, the strain becomes increasingly more compressive (negative values). An interesting sharp tensile strain (indicated by the arrows) can be observed exactly at the $TiO_2$-$VO_2$ interface, induced by the lattice mismatch. Nevertheless, some local fluctuations of strain in the atomic layers are also noticeable in the out-of-plane direction in the $VO_2$ film. The fluctuating strain values predominantly correlate with the M2 polymorph domain (in Zone III). This seems to be in line with the observed strain behavior reported by Rodriguez et al. [32] that can be attributed to the formation of the intermediate phases in $VO_2$ films. Due to Poisson's ratio, obviously a tensile strain must be present along the in-plane direction [110] of the film as Fig. 4 suggests. Thus, the stabilization of the M2 phase at room temperature can be attributed to the prevalent strain gradient generated within the $VO_2$ thin film on the $TiO_2$ substrate upon cooling from high temperature.

In addition to the studied structural aspects, Figs. 4C-D demonstrate the electron energy loss spectroscopy (EELS) analysis across the film to understand the local electronic properties and the band gap characteristics.

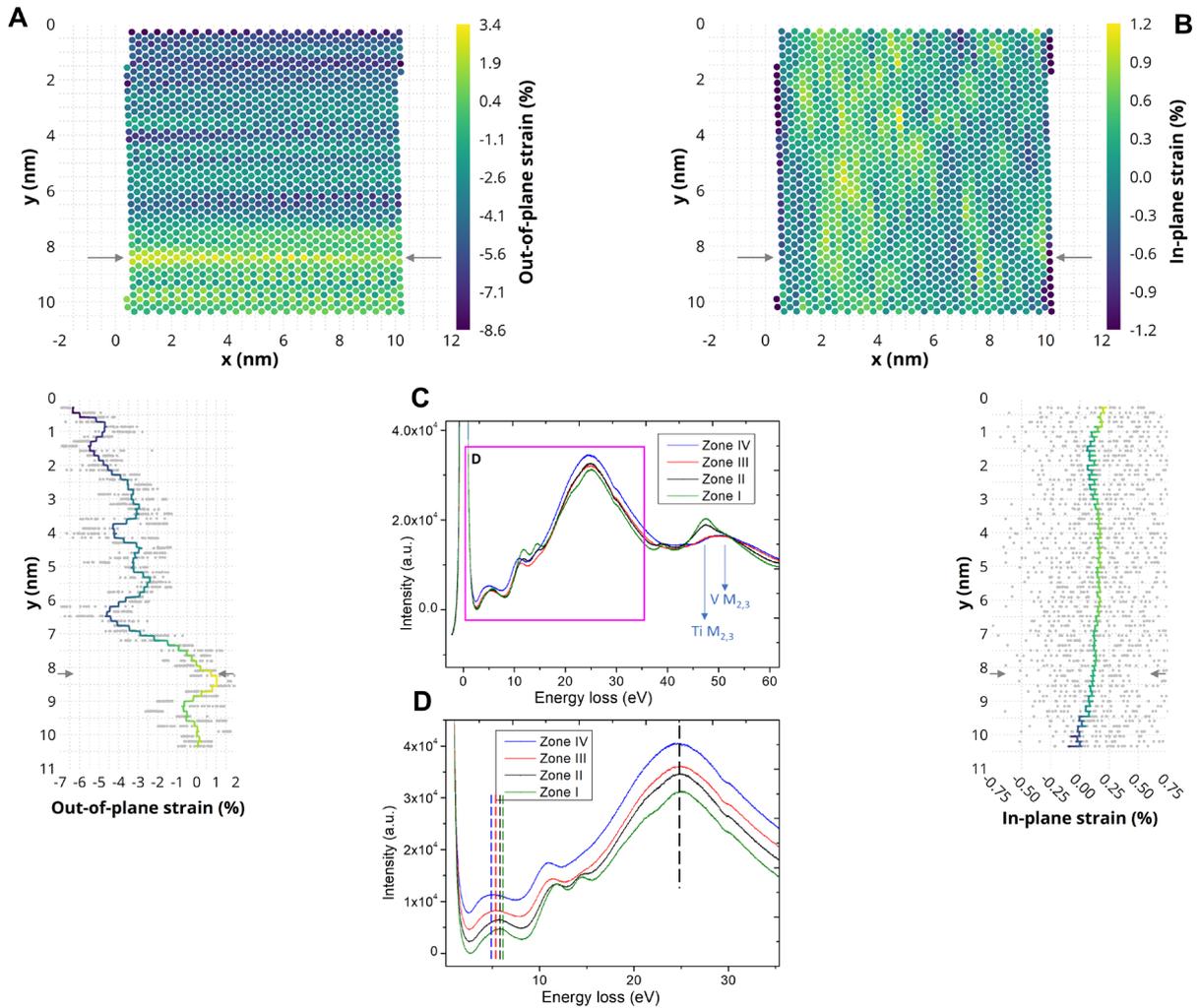

**Fig. 4. Atomic-structure strain mapping and EELS analysis.**

(A) Strain analysis along the growth direction $[001]_R$ (out-of-plane) showing the compressive strain values in $VO_2$ film (the arrows indicate the interface between the $TiO_2$ substrate and $VO_2$ film), (B) strain analysis along the in-plane direction $[110]_R$ of the film, revealing a tensile strain generated in the $VO_2$ film, (C) EELS spectra recorded from cross-section lamella of $VO_2$ on $TiO_2$ (001) with projection [100] near zero-loss peak, (D) higher resolution of the spectra from zero loss peak to the plasmon region, with equal vertical offsets for clarity.

Electron energy loss spectra were recorded in the low-loss region, from the zero-loss peak to beyond the Ti and V $M_{2,3}$ absorption edges as shown in Fig. 4C-D. Spectra were recorded in the four different Zones I to IV as indicated in Fig. 2c in different parts of the sample. The Ti and V $M_{2,3}$ edges are important to confirm that Zone II is indeed an intermixed region, because the intensity of the Ti $M_{2,3}$ is appreciable, but that Zone III is effectively the same as Zone IV which must be rather pure $VO_2$. The peaks in the range of 0-30 eV before the $M_{2,3}$ edges arise from the collective excitations from the valence bands to the conduction bands of the respective sample layers. When (the maxima in) the zero loss peaks are aligned well, the maximum intensities in the main plasmon peak at about 25 eV are also aligned well for all four zones as indicated by the vertical dashed black line in Fig. 4D. However, systematic differences are observed in the position of the first peak after the zero-loss peak at about 5 eV that can be associated with states close to the band gap ($E_g$). Indeed, in all different parts of the films where we recorded spectra always a systematic trend was observed that this peak moved to increasingly lower energies when going from Zone I via II and III to Zone IV. This suggests that the effective $E_g$ reduces systematically when going from Zone I to IV. However, a more accurate method to determine the value of the band gap is to extrapolate the linear fit to the slope of the low energy side of this peak towards lower energy and then the intercept of this linear fit with the horizontal axis is a measure of the band gap. The exact procedure we performed to do this is described in detail in the SI.

Quantifying the band-gap using this procedure, it was found to be 2.89 eV for pure $TiO_2$ substrate (Zone I), 2.65 eV in Zone II (intermixed layer), 2.43 eV in Zone III (M2 containing layer) and finally 1.92 eV in the M1 region (Zone IV). The value we measured for rutile $TiO_2$ 2.9 eV is rather close to its known value of 3.0-3.1 eV [42,43]. The value for the M1 phase of $VO_2$ 1.9 eV is somewhat larger than reported in some earlier works [44,45]. Note that in the different parts of the sample analyzed the $E_g$ values we derived differ by about 0.2-0.3 eV, but that the trend was always very systematic, i.e. $E_g$ of Zone I > Zone II > Zone III > Zone IV. This shows that a significant change in the band-gap occurs for the layers and also allows us to differentiate between the Zones III and IV and thus between the M1 and M2 monoclinic layers of the sample. Note that Zone IV can be considered pure M1 phase and that in Zone III the M2 phase is present, but not exclusively since also some M1 occurs also due to the overlap that occurs in the projection direction in the relative thick TEM sample as produced by FIB. The actual difference in band gap between the M2 and M1 phase in our sample is therefore larger than the values we measured here for the difference in $E_g$ between Zones III and IV.

**Conclusions**

In summary, we have studied the cross-section of epitaxially grown $VO_2$ films on $TiO_2$ (001) using atomic-resolution STEM. Using iDPC, we have directly resolved both V and O atomic columns simultaneously and observed the presence of the intermediate monoclinic M2 phase which is stabilized by strain in thin nanolayers (~2 nm) above the interface. We have distinguished this M2 phase from the main monoclinic M1 phase using a set of criteria involving the symmetry of bonds and ellipticity of atoms (both cations and anions). The methodology developed here based on state-of-the-art STEM images in which also the oxygen atomic columns are resolved accurately (like possible with iDPC images), and in which also the ellipticity of projected atomic columns can be quantified, can be applied to a wide range of oxides or materials containing a mixture of low and high Z atomic columns. This is particularly important for improving our understanding of the atomic structures of materials varying at nanometer length scales that cannot be captured with methods that analyze materials on a more global scale like diffraction based techniques.

The comprehensive detection of the different zones in the $VO_2$ film (especially the M2 domain) can potentially have a significant impact on the design of modern optoelectronics and memristors. Being able to resolve the spatial configuration of the M2 (and M1) phases, provides the prospect to tune the amount and distribution of such intermediate phases in order to optimize the electronic properties within the $VO_2$ films.

**Acknowledgment:**

M.H. and A.M. acknowledge financial support by the Deutsche Forschungsgemeinschaft (DFG) in the project HE 7192/7-1. We acknowledge the support of Hitachi High-Technologie. EvdV acknowledges financial support from the Groningen Cognitive Systems and Materials Center (CogniGron) and the Ubbo Emmius Foundation.